%% file: draft_CL1.tex
\documentclass[a4paper,11pt]{article}
\pdfoutput=1 
\usepackage{jheppub} 

\usepackage[T1]{fontenc} 

\RequirePackage[displaymath]{lineno} 
\usepackage{epstopdf}
\usepackage{epsfig}
\usepackage{graphicx}
\usepackage{dcolumn}
\usepackage{bm}
\usepackage{ltablex,booktabs}
\usepackage{overpic}
\usepackage{subfigure}
\usepackage{float}
\usepackage{color}

\usepackage{amsmath}
\usepackage{mathcomp}
\usepackage{mathrsfs}
\usepackage{multirow}
\usepackage{rotating}
\usepackage{amssymb}
\usepackage{gensymb}
\usepackage{amsmath}
\usepackage{tabularx}
\usepackage{threeparttable}
\usepackage{caption}
\DeclareCaptionLabelFormat{up}{\MakeUppercase{#1} #2}
\usepackage{enumitem}
\usepackage[labelsep=period]{caption}
\DeclareCaptionLabelFormat{figure}{\textbf{FIG}~\textbf{\arabic{figure}}}
\title{Study of the $e^+e^-\to \pi^+\pi^-D_s^+D_s^-$ process from $\sqrt{s}$ = 4.42 to 4.95~GeV at BESIII}

\collaborationImg{\includegraphics[width=0.15\textwidth, angle=90]{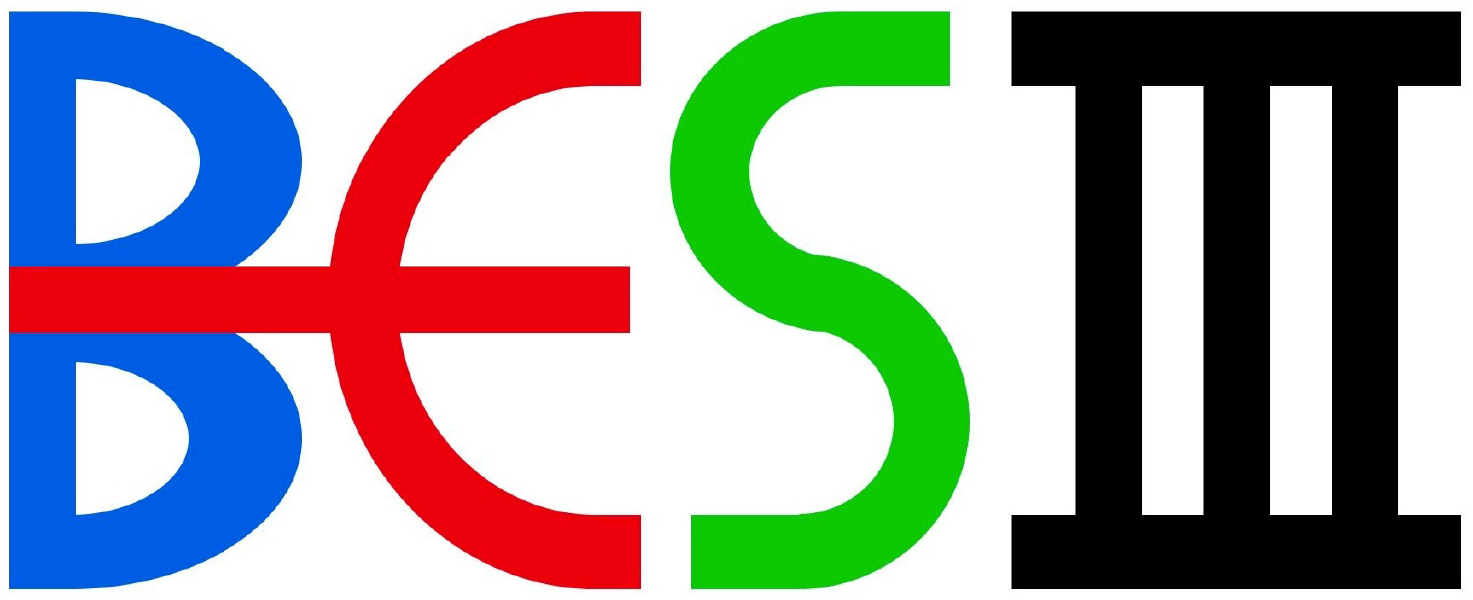}}
\collaboration{BESIII Collaboration}

\author{\input{author}}

\abstract{
Based on $8.5~\text{fb}^{-1}$ of
$e^+e^-$ collision data collected at center-of-mass  energies between 4.42 and 4.95~GeV with the BESIII detector at the BEPCII storage ring, we investigate the process $e^+e^-\to \pi^+\pi^-D_s^+D_s^-$. With no significant signal observed, upper limits on the Born cross sections of $e^+e^-\to \pi^+\pi^-D_s^+D_s^-$ at each energy value at the 90\% confidence level are determined. Additionally, a search for intermediate charmonium-like resonances is performed in the $M(D_s^+ D_s^-)$ invariant mass spectrum, but no significant resonant structures are observed with the current statistics.
}
\arxivnumber{}
\let\oldequation\equation
\let\oldendequation\endequation

\renewenvironment{equation}
  {\linenomath\oldequation}
  {\oldendequation\endlinenomath}

\let\oldeqnarray\eqnarray
\let\oldendeqnarray\endeqnarray

\renewenvironment{eqnarray}
  {\linenomath\oldeqnarray}
  {\oldendeqnarray\endlinenomath}
\begin{document}
\newcommand{\BESIIIorcid}[1]{\href{https://orcid.org/#1}{\hspace*{0.1em}\raisebox{-0.45ex}{\includegraphics[width=1em]{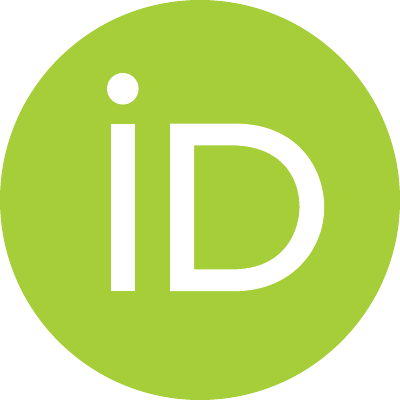}}}}
\maketitle
\flushbottom

\section{Introduction}
$~~~~$
In recent years, the exploration of charmonium-like states has become a prominent focus in hadron physics. Vector charmonium-like states with masses above the open-charm threshold, as for example  $Y(4230)$, $Y(4360)$ and $Y(4660)$, have been observed in many experiments, including BABAR~\cite{BABAR1,BABAR2,BABAR3,BABAR4}, Belle~\cite{Belle,Belle2,Belle3,Belle4}, CLEO~\cite{CLEO,CLEO1}, and BESIII~\cite{BES,BES2,BES3}. A thorough investigation of these states in the open-charm channel can provide crucial insights into their characteristics and test different theoretical explanations such as hybrid charmonium~\cite{hybrid}, tetraquark~\cite{tetraquark}, and hadronic molecule~\cite{hadronic,Wang:2025dur} models. 

The BESIII experiment has identified various charmonium-like states in open-charm processes, including $Y(4220)$, $Y(4390)$~\cite{PRL122-102002}, $Z_{c}$(3900)~\cite{PRL112-022001}. In 2021, BESIII announced the observation of a novel exotic multi-quark state candidate, $Z_{cs}(3985)^-$, produced in the reaction $e^+e^-\to K^+(D_s^-D^{*0} + D_s^{*-}D^0)$; the production occurred at the center-of-mass~(c.m.) energy ($\sqrt{s}$) of 4.68~GeV~\cite{3985}.
The revelation of $Z_{cs}(3985)^-$ has stimulated a fresh wave of investigations into multi-quark structures. Researchers are particularly focusing on the interplay between charm and strange quarks within these states. 
Recently, the LHCb experiment unveiled a distinct and intriguing hadronic state in the $D_s^+D_s^-$ system originating from the three-body decay $B^+\to D_s^+D_s^-K^+$, which indicates the potential presence of a four-quark ($c\overline{c}$$s\overline{s}$) element ~\cite{lhcb}. 

In this analysis, we investigate the process \(e^+e^-\to\pi^+\pi^-D_s^+D_s^-\), focusing on the interactions of the two $D_s^\pm$ mesons, which are charm-strange quark bound states. By analyzing data from the BESIII detector at c.m. energies ranging from 4.42 to 4.95~{\rm GeV}~\cite{lum1,lum2}, we measure the Born cross sections. This study aims to enhance our understanding of charm and strange quark dynamics in multi-quark systems, and provide insights into the structures composed of charm and strange quarks~\cite{Prelovsek:2020oig,Liu:2021fba}.

\section{Detector and Data Samples}
\label{sec:detector_dataset}
$~~~~$The BESIII detector~\cite{bes3-dector} records symmetric $e^+e^-$ collisions provided by the BEPCII storage ring~\cite{detector2} in the c.m. energies range from 1.84 to 4.95~GeV, with a peak luminosity of~$1.1 \times 10^{33}\;\text{cm}^{-2}\text{s}^{-1}$ 
achieved at $\sqrt{s} = 3.773\;\text{GeV}$. BESIII has collected large data samples in this energy region~\cite{det3,det4,det5}. 

The cylindrical core of the BESIII detector covers 93\% of the full solid angle and consists of a helium-based multilayer drift chamber~(MDC), a time-of-flight system~(TOF), and a CsI(Tl) electromagnetic calorimeter~(EMC), which are all enclosed in a superconducting solenoidal magnet providing a 1.0~T magnetic field. The solenoid is supported by an octagonal flux-return yoke with resistive plate counter muon identification modules interleaved with steel. The charged-particle momentum resolution at $1~{\rm GeV}/c$ is $0.5\%$, and the ${\rm d}E/{\rm d}x$ resolution is $6\%$ for electrons
from Bhabha scattering. The EMC measures photon energies with a resolution of $2.5\%$ ($5\%$) at $1$~GeV in the barrel (end cap) region. The time resolution of the TOF barrel part is 68~ps, and the end-cap TOF system was upgraded in 2015 with a multi-gap resistive plate chamber technology, providing a time resolution of 60~ps, which benefits 73\% of the data used in this analysis~\cite{det6,det7,det8}.

The analysis is performed within the framework of the BESIII Offline Software System (BOSS)~\cite{BOSS} which handles detector calibration, event reconstruction and data storage. The dataset comprises data samples at 15 energy values. 
The integrated luminosities are determined through large-angle Bhabha events, and the c.m. energies of the data samples are obtained from offline measurements~\cite{lum2}. 
Monte Carlo (MC) simulations are employed to determine detection efficiencies and estimate backgrounds. They are generated with a {\sc
geant4}-based~\cite{GEANT4-Col} software package, which
includes the geometric description of the BESIII detector and the detector response, are used to determine detection efficiencies and to estimate backgrounds. The simulation models the beam energy spread and initial state radiation (ISR) in $e^+e^-$ annihilations using the {\sc
kkmc} generator~\cite{kkmc}. Inclusive MC samples for background assessments are generated in a range of $\sqrt{s}$ = 4.42 to 4.95~GeV. All particle decays are modelled with {\sc
evtgen}~\cite{evtgen,simulation-RGPing} using branching fractions either taken from the Particle Data Group (PDG)~\cite{pdg}, when available,
or otherwise estimated with {\sc lundcharm}~\cite{PRD62-034003}.
      
      The signal processes involve the $e^+e^-\to \pi^+\pi^-D_s^+D_s^-$ reaction, where the decays are generated by BesEvtGen, including $D_s^+\to\ K_S^0K^+$  and  $K^+K^- \pi^+$, $D_s^-\to\ K_S^0K^-$  and  $K^+K^- \pi^-$. The decays of $D_s^+\to\ K_S^0K^+$ and $D_s^-\to\ K_S^0K^-$ follow the phase space model (PHSP), while $D_s^+\to\ K^+K^- \pi^+$ and $D_s^-\to\ K^+K^- \pi^-$ adhere to the D\_DALITZ~\cite{DALITZ1,simulation-RGPing} model.

\section{Event Selections and Analysis Techniques}
\label{chap:event_selection}

$~~~~$A full reconstruction method is used to extract the signal process $e^+e^- \to \pi^+\pi^-D_s^+D_s^-$, where the $D_s^\pm$ mesons decay into $K^+K^-\pi^\pm \text{ or } K^0_S(\to \pi^+\pi^-)K^\pm$. Charged tracks are reconstructed from MDC hits within a polar-angle ($\theta$) acceptance range of $|\cos\theta| < 0.93$, where $\theta$ is defined with respect to the symmetry axis of the MDC (defined as the $z$-axis). For the charged tracks not originating from $K^0_S$ decays, the distance of the closest approach to the interaction point (IP) must be less than 10 cm along the $z$-axis ($V_z$), and less than 1~cm in the transverse plane ($V_{ xy}$). The ${\rm d}E/{\rm d}x$ and TOF information for each charged track are combined to calculate the Particle Identification (PID) probabilities $P_i$ ($i = \pi, K$) for each particle-type hypothesis. Kaon candidates are required to have $P_K > P_\pi$, while pion candidates must satisfy $P_\pi > P_K$. 

Each $K^0_S$ candidate is reconstructed from two oppositely charged tracks satisfying the conditions $|V_z| < 20 \ \text{cm}$ and $|\cos\theta| < 0.93$. The two charged tracks are assigned as $\pi^+\pi^-$ without imposing PID criteria. They are constrained to originate from a common vertex and are required to have an invariant mass ($M_{\pi^+\pi^-}$) within the interval (0.487, 0.511)~GeV/$c^2$. The decay length of the $K^0_S$ candidate is required to be greater than twice the vertex resolution.

Selected $K^{\pm}$, $\pi^{\pm}$, $K^0_S$ candidates are used to reconstruct the $D_s^{\pm}$ meson candidates. For events with multiple candidate combinations, the optimal pair is selected by 
minimizing the variable $\Delta \hat{M} = |\hat{M} - M_{\text{PDG}}(D_s^{\pm})|$, 
where $\hat{M} = [M(D_s^+) + M(D_s^-)]/2$ is the average reconstructed mass of 
the $D_s^+ D_s^-$ pair, and $M_{\text{PDG}}(D_s^{\pm})$ is the nominal mass 
from the PDG~\cite{pdg}.

\begin{figure*}[htbp] 
\centering
\begin{minipage}[b]{0.48\textwidth}
\centering
\includegraphics[width=1.0\linewidth]{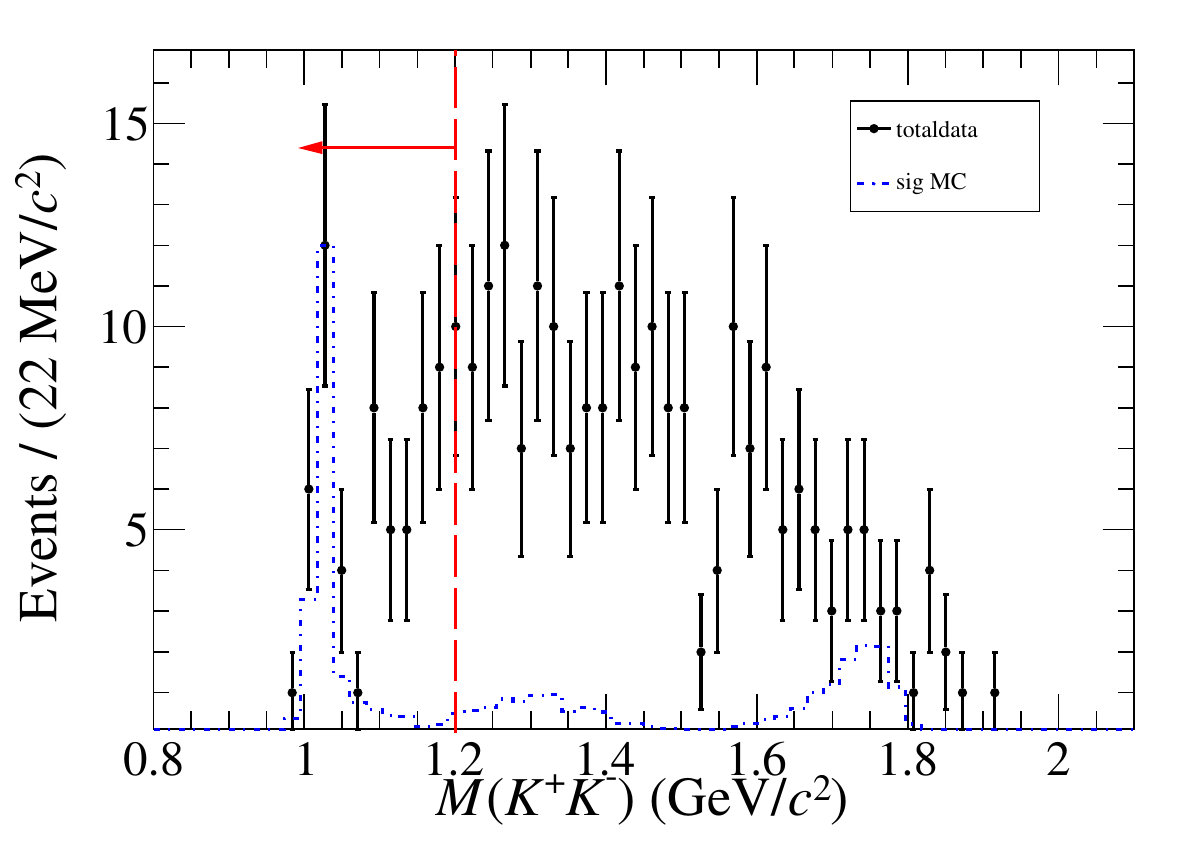} 
\put(-25,115){(a)} 
\end{minipage}
\hfill
\begin{minipage}[b]{0.48\textwidth}
\centering
\includegraphics[width=1.0\linewidth]{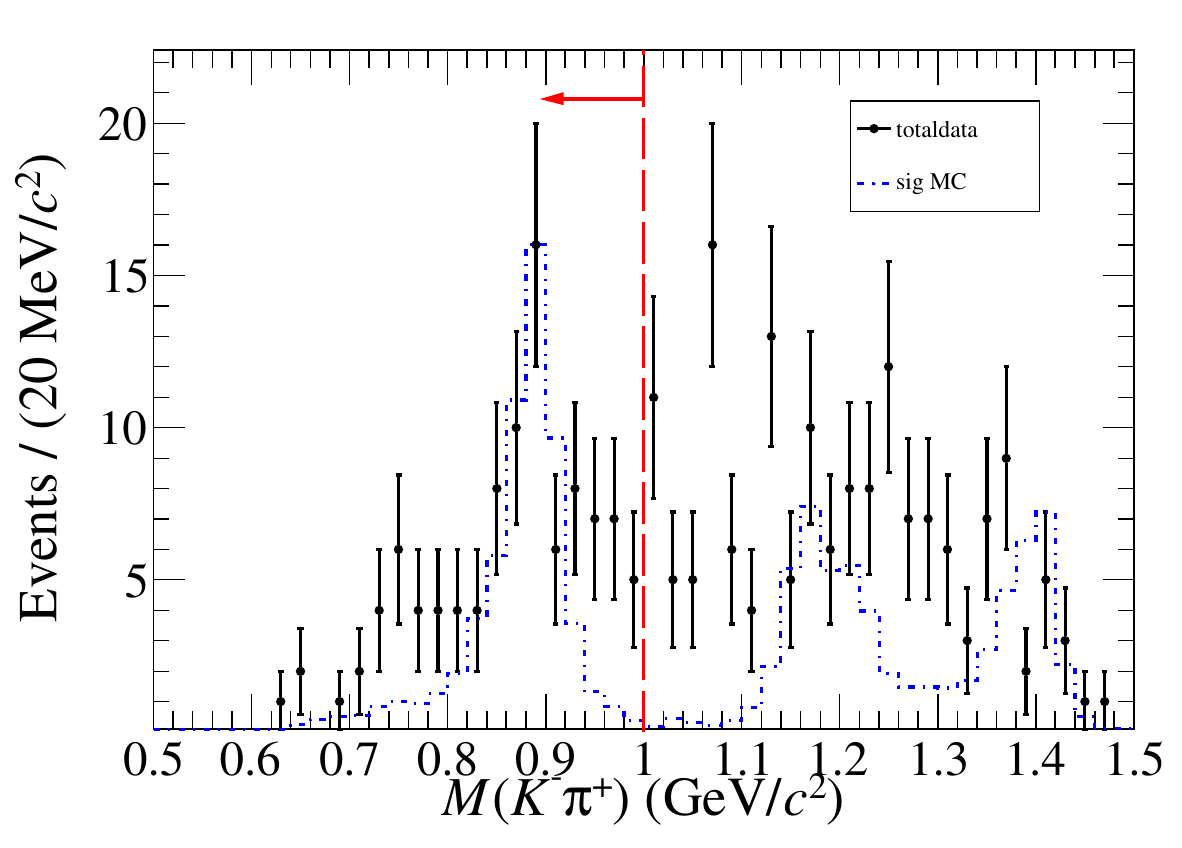}
\put(-25,115){(b)}
\end{minipage}
\caption{The invariant mass of $M(K^+K^-)$ (a) and $M(K^-\pi^+)$ (b). The black dots with error bars represent the data for fifteen energy points, the solid lines (blue) describe the signal MC at $\sqrt{s}$ = 4.42 GeV.}
\label{kkkpi}
\end{figure*}

In order to suppress backgrounds, mass window requirements for the intermediate states $\phi$ and $\overline{K}^{*0}$ are applied during the $D_s^+ \to K^-K^+\pi^+$ candidate selection. Based on the observed invariant mass distributions of  $M(K^+K^-)$ and $M(K^-\pi^+)$ as shown in Fig.~\ref{kkkpi}, the mass window for the $\phi$ in the $D_s^+ \to \phi \pi^+$ decay is required to be $M(K^+K^-) < 1.2$~GeV/$c^2$. Similarly, for the $\overline{K}^{*0}$ in $D_s^+ \to \overline{K}^{*0} K^+$, the mass window is set as $M(K^-\pi^+) < 1.0$~GeV/$c^2$. Corresponding selection criteria are also applied to the~$D_s^-$ candidates.

The inclusive MC samples are utilized to examine potential background contributions. After applying all the specified requirements, a total of 73 events remain. All of them are non-dominant backgrounds, and no significant peaking background is observed within the $D_s^{\pm}$ signal region.

\section{Signal Extraction}
\begin{figure*}[htbp] 
\centering
\begin{minipage}[b]{0.48\textwidth}
\centering
\includegraphics[width=1.0\linewidth]{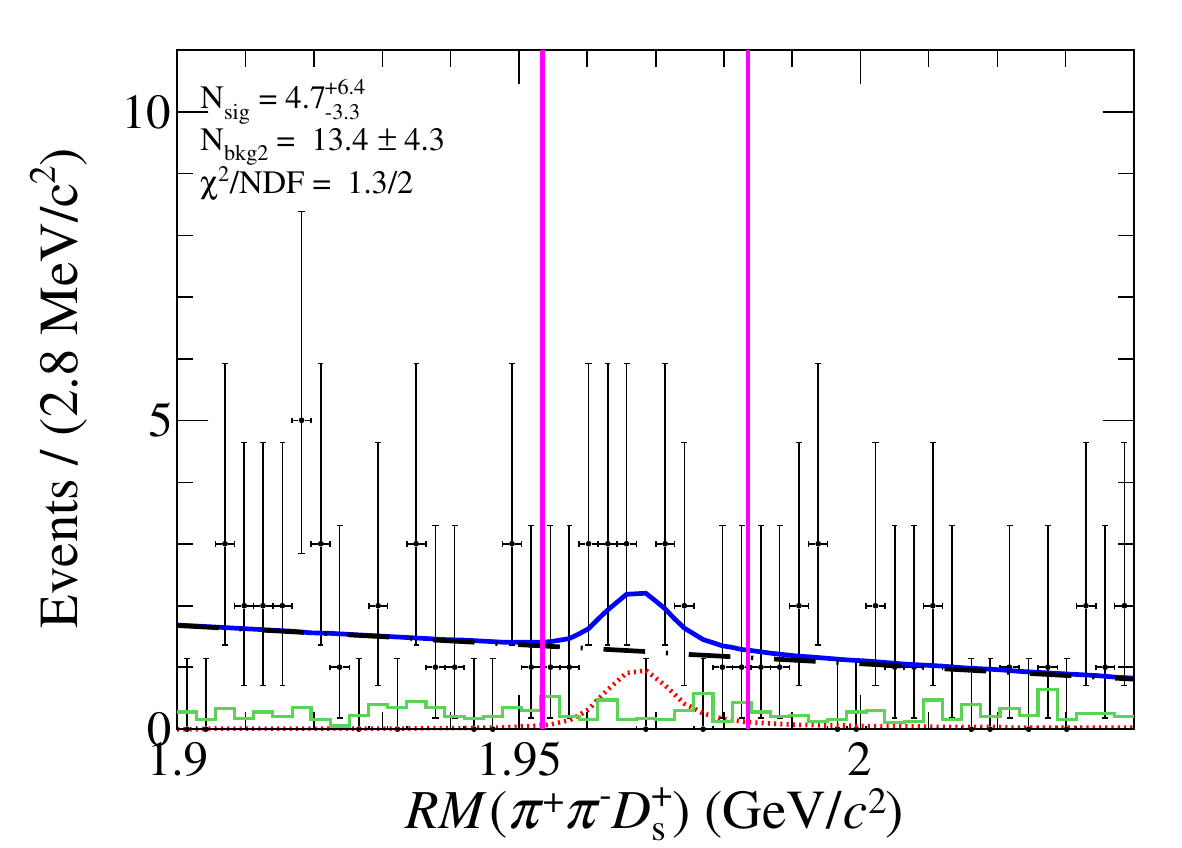} 
\put(-25,115){(a)} 
\end{minipage}
\hfill
\begin{minipage}[b]{0.48\textwidth}
\centering
\includegraphics[width=1.0\linewidth]{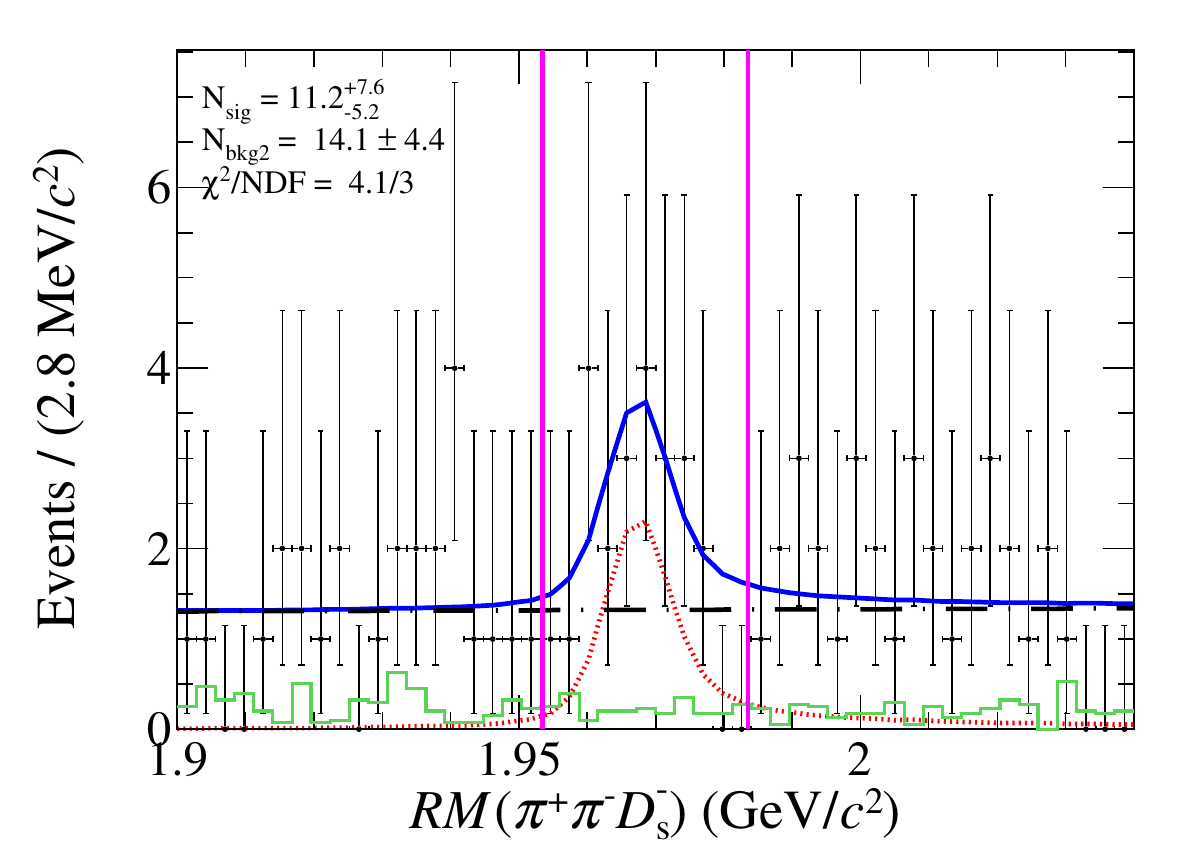}
\put(-25,115){(b)}
\end{minipage}
\caption{Fits to the $RM(\pi^+\pi^-D_s^+)$ and $RM(\pi^+\pi^-D_s^-)$ distributions in (a) and (b), respectively. The black dots with error bars represent data, the solid blue lines show the total fits, the dashed red lines indicate the signal shapes, the green line shows the background from inclusive MC after scaling, and the dash-dotted black lines represent the background.}
\label{fit_mDst_4}
\end{figure*}

$~~~~$The signal yield at each energy point is obtained by performing a 
one-dimensional (1D) unbinned maximum likelihood fit to the 
corresponding $RM(\pi^+\pi^-D_s^{\pm})$ distribution. The mass window requirement $|M(D_s^-) - M_{\rm PDG}(D_s^-)| < 15~{\rm MeV}/c^2$ is applied 
in the fit to the $RM(\pi^+\pi^-D_s^-)$ distribution,
and similarly for the $RM(\pi^+\pi^-D_s^+)$ fit. To improve the mass resolution, the corrected recoil mass $RM(\pi^{+}\pi^{-}D_{s}^{+}) + M(D_{s}^{+}) - M_{\rm PDG}(D_{s}^{+})$ is used. The signal shape is modeled by a non-parametric PDF, which is a histogram-based PDF that describes the signal shape by interpolating the distribution of the simulated MC events. Meanwhile, a first-order polynomial function models the background shape. In the fitting process, both the signal and background yields, as well as the polynomial function parameters for the background shape, are floated. The data from the fifteen samples are combined to enhance the statistical significance. The fits to the $RM(\pi^+\pi^-D_s^+)$ and~$RM(\pi^+\pi^-D_s^-)$  distributions are illustrated in Fig.~\ref{fit_mDst_4}. The $\chi^2$/NDF values are 1.3/2 for~$RM(\pi^+\pi^-D_s^+)$ and 4.1/3 for $RM(\pi^+\pi^-D_s^-)$. The signal yields in the $RM(\pi^+\pi^-D_s^+)$ and $RM(\pi^+\pi^-D_s^-)$ fits are $4.7^{+6.4}_{-3.3}$ and $11.2^{+7.6}_{-5.2}$, respectively. The statistical significance of the signal is determined to be 1.2$\sigma$ and 2.6$\sigma$ by comparing the likelihood values with and without the signal assumption in the fits, considering the change in degrees of freedom. For a conservative estimation of the significance, the outcome from the fit to $RM(\pi^+\pi^-D_s^+)$ is chosen as the nominal result.

\section{Search for the Charmonium-like States in the $D_s^+D_s^-$ System}

\begin{figure}[htbp] 
\centering
\includegraphics[width=0.6\linewidth]{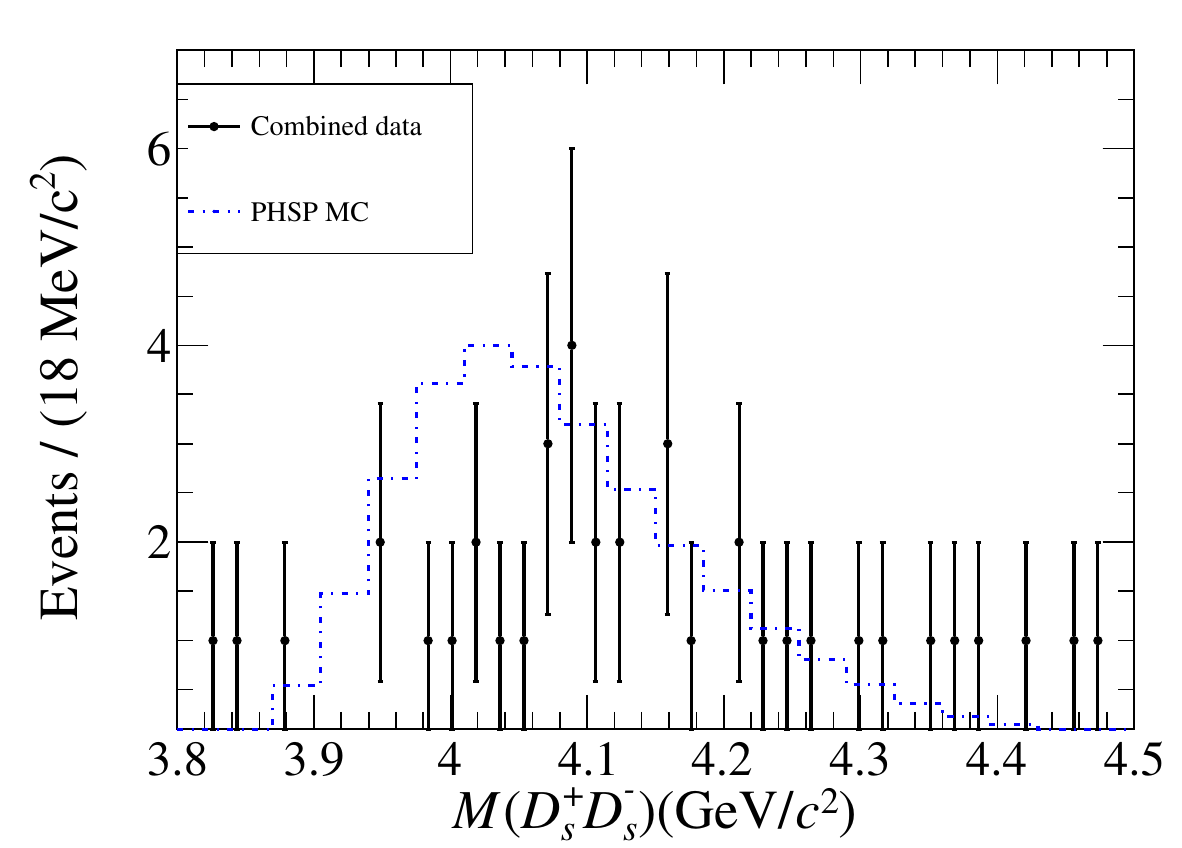} 

\vspace{-8pt} 

\caption{The $M(D_s^+D_s^-)$ invariant mass distribution for the combined data samples. The black dots with error bars represent data, and the dotted blue curve shows the weighted PHSP component.}

\label{dsds}
\end{figure}

$~~~~$To explore potential charmonium-like intermediate states, the $M(D_s^+D_s^-)$ invariant mass spectrum is examined using the combined data samples from all fifteen energy values, corresponding to a total integrated luminosity of 8.5~fb$^{-1}$.
As shown in Fig.~\ref{dsds}, the $M(D_s^+D_s^-)$ distribution is compared with the PHSP MC component. The PHSP MC shape (blue dashed line) is obtained by combining the PHSP MC samples from each energy value, weighted by their respective measured Born cross sections, selection efficiencies, and integrated luminosities. To reduce statistical fluctuations, the weighted PHSP MC distribution is further smoothed.

No significant resonant structures are observed in the $M(D_s^+ D_s^-)$ system with the current data sample, and the result remains statistically limited.

\section{Born Cross Section Measurement}

\begin{figure}[htbp] 
\centering
\includegraphics[width=1.0\linewidth]{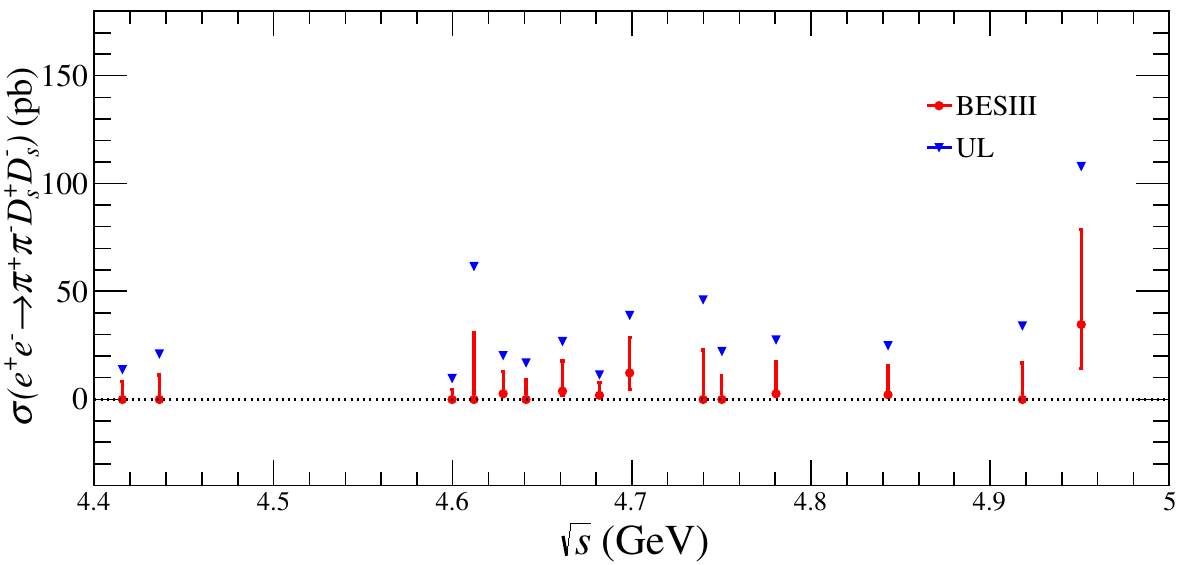} 

\vspace{-8pt} 

\caption{Born cross sections of $e^+e^-\to \pi^+\pi^-D_s^+D_s^-$ at different center-of-mass energies. The red dots with error bars represent the measurements from this analysis, while the blue triangles indicate the upper limits.}

\vspace{-5pt} 
\label{dressed}
\end{figure}

$~~~~$The Born cross section for the process $e^+e^-\to \pi^+\pi^-D_s^+D_s^-$ can be calculated as follows
\begin{linenomath*}
\begin{eqnarray}
\begin{aligned}
\sigma(e^+e^-\to\pi^+\pi^-D_s^+D_s^-)\
= \frac{N^{\rm sig}}{\mathcal{L}(1+\delta^{r})(1+\delta^{v})
\sum_{\substack {i,j}}\epsilon_{ij}\mathcal{B}_{i}\mathcal{B}_{j}},
\label{eq111}
\end{aligned}
\end{eqnarray}
\end{linenomath*}
where $N^{\rm sig}$ is the signal yield, $\mathcal{L}$ is the integrated luminosity~\cite{lum2},
$\epsilon_{ij}$ indicates the signal efficiency for $e^+e^-\to \pi^+\pi^-D_s^+D_s^-$ with $D_s^+\to i$ and $D_s^-\to j$; $\mathcal{B}_{i}$ and $\mathcal{B}_{j}$ are the branching fractions for $D_s^+\to i$ and $D_s^-\to j$, respectively; $1+\delta^{r}$ is the ISR correction factor, which is defined as  
\begin{linenomath*}
\begin{eqnarray}
\begin{aligned}
1+\delta^r=\frac{\sigma^{\rm obs}}{\sigma^B}=\frac{\int\sigma^B(s(1-x))F(x,s)dx}{\sigma^B(s)},
\label{eq111}
\end{aligned}
\end{eqnarray}
\end{linenomath*}
where $F(x,s)$ is the radiator function derived from QED calculations~\cite{EAVS} with 0.1\% accuracy. The input line shape $\sigma^B$ in Eq.~(\ref{eq111}) is parameterized by a cubic 
polynomial function fitted to the initially measured Born cross sections. 
This line shape is then used to determine the final ISR correction factors and efficiencies for all data samples. The factor $1+\delta^{v}$ is the vacuum polarization (VP) correction factor calculated using the Fortran package provided by Fred Jegerlehner~\cite{EPJC-66-585}, which determines the correction by summing the leptonic contribution calculated in precise perturbative QED and the non-perturbative hadronic contribution evaluated via dispersion relations using experimental data for electron-positron annihilation into hadrons.
Figure~\ref{dressed} shows the measured Born cross sections and their upper limits obtained from this analysis (blue dots).

\section{Systematic Uncertainties}
\label{chap:SYSTEMATICS}
$~~~~$Systematic uncertainties are mainly from the luminosity measurement, the quoted branching fractions, the ISR correction factor, and the fit schemes. Due to the limited statistics and large fluctuations at individual energy points, the systematic uncertainties are evaluated using the combined data sample when applicable, in order to obtain a stable estimate. 
For this purpose, the average Born cross section is calculated as
\begin{linenomath*}
\begin{eqnarray}
\begin{aligned}
\bar{\sigma}(e^+e^- \to \pi^+\pi^- D_s^+ D_s^-)
= \frac{N_{\rm tot}}{\displaystyle\sum_{k} \left[ \mathcal{L}_k (1+\delta^r_k)(1+\delta^v_k) \sum_{i,j} \epsilon_{ij}^{(k)} \mathcal{B}_i \mathcal{B}_j \right]},
\label{eq112}
\end{aligned}
\end{eqnarray} 
\end{linenomath*}
where $N_{\rm tot}$ is the total signal yield across all fifteen energy values, $\mathcal{L}_k$ is the integrated luminosity of the $k$-th energy value, $1+\delta^r_k$ and $1+\delta^v_k$ are the ISR correction factor and the VP correction factor at the $k$-th energy value, respectively.
A summary of the systematic uncertainties is given in TABLE~\ref{tab:sys}. The total systematic uncertainty is obtained by summing the individual contributions in quadrature.

The integrated luminosities of these data samples are measured using Bhabha events, with an uncertainty of 1.0\%~\cite{CPC39-093001}.

The systematic uncertainty due to the charged $\pi$ and $K$ tracking from the interaction point (IP) is estimated to be 1.0\% per track~\cite{A,B,C}. 
The detection efficiencies are varied within one standard deviation according to the charged track multiplicities in the $D_s^{\pm}$ reconstruction channels; 
the largest difference in the cross sections, 7.0\%, is taken as the systematic uncertainty. 
Similarly, the systematic uncertainty due to PID is taken to be 1.0\% per charged track~\cite{A,B,C}. 
The systematic uncertainty associated with $K^0_S$ reconstruction is taken as 0.4\%~\cite{yangyingmemo}.

The branching fractions of $D_{s}^{+} \to K_{S}^{0} K^{+}$ and $D_{s}^{+} \to K^{+} K^{-} \pi^{+}$ are taken from the PDG~\cite{pdg}. Their uncertainties are propagated to the final cross section measurement by varying the branching fractions by $\pm 1\sigma$ in the calculation. The resulting largest differences from the nominal cross sections are assigned as the systematic uncertainties, which are determined to be 0.5\% and 2.2\%, respectively.

\begin{figure*}[htbp] 
\centering
\begin{minipage}[b]{1.0\linewidth} 
\centering
\includegraphics[width=1.0\linewidth]{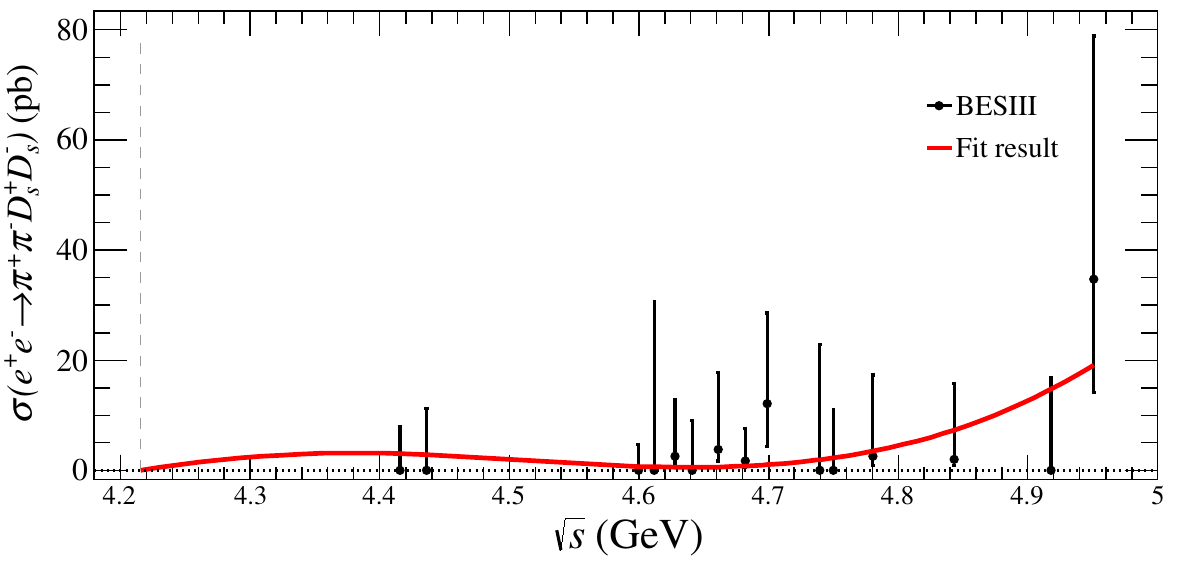} 
\put(-360,175){(a)} 
\end{minipage}

\vspace{-5pt} 

\begin{minipage}[b]{1.0\linewidth}
\centering
\includegraphics[width=1.0\linewidth]{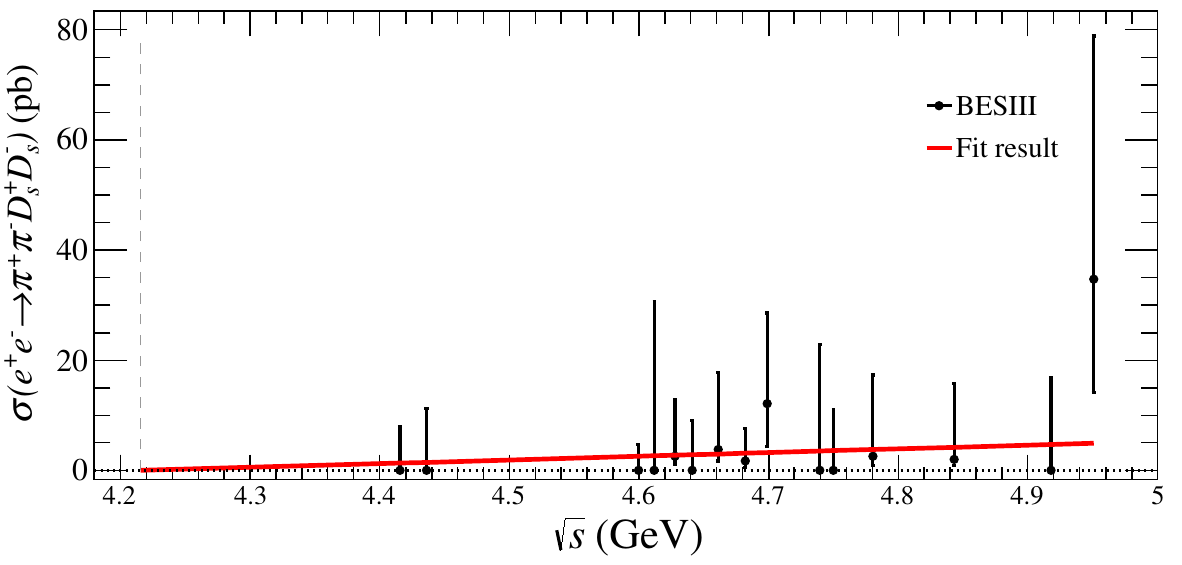}
\put(-360,175){(b)}
\end{minipage}

\caption{Born cross section line shapes for $e^+e^- \to \pi^+\pi^-D_s^+D_s^-$ fitted with (a) a cubic polynomial and (b) a linear polynomial.}
\label{weight}
\end{figure*}

The uncertainty in the line shape of the cross sections can introduce systematic uncertainties in both the ISR correction factor and the signal efficiency. To account for this, the measured cross sections are fitted with a cubic polynomial to obtain an updated line shape. This updated shape is then incorporated into the calculation of the ISR correction factors and efficiencies. The systematic uncertainty is assessed by comparing these results with those obtained using an alternative linear polynomial model, as shown in Fig.~\ref{weight}. The difference on the average cross section, 1.5\%  is taken as the systematic uncertainty.

The $D_s^\pm$ mass resolution discrepancy between data and MC simulation affects the efficiencies. The uncertainty due to the mass window for the $D_s^{\pm} \to K_S^0 K^{\pm}$ decay is assessed by varying the nominal mass window by $\pm 15$~MeV/$c^2$. The relative difference between the observed cross sections and the nominal results, 2.2\%, is taken as the systematic uncertainty. 
For the $D_s^{\pm} \to K^+ K^- \pi^{\pm}$ decay, the systematic uncertainty is taken to be 1\% according to Ref.~\cite{ISR}.

The systematic uncertainty associated with the signal yield encompasses uncertainties in the fitting range, the signal shape, and the background shape. The uncertainty related to the fitting range is evaluated using the Barlow test~\cite{zeta}. By varying the fit range, the changes in the final results consistently remain below \(2\sigma\), making this uncertainty negligible. To estimate the systematic uncertainty related to the $RM(\pi^+\pi^-D_s^\pm)$ signal shape, we account for the resolution difference between data and MC simulation. The nominal MC-derived shape is convolved with a Gaussian function~\cite{liutong}. The resulting difference in the signal yields, 6.4\%, is taken as the systematic uncertainty. For the background shape, the uncertainty is estimated by replacing the first-order polynomial with a second-order polynomial, and the resulting difference in fitted signal yields is less than 0.1\%, which is negligible.

\begin{table}[h]
\begin{center}
\caption{\small Systematic uncertainties for the Born cross section (\%).}
\begin{tabular}{c|c}
\hline
Source   & Uncertainty (\%)   \\
\hline
Integrated luminosity            &     1.0  \\
Tracking                         &     7.0 \\
PID                              &     7.0 \\
$K_S^0$ reconstruction             &     0.4\\
$\mathcal{B}(D_s^+\to K_S^0K^+)$ &     0.6 \\
$\mathcal{B}(D_s^+\to K^+K^-\pi^+)$&     2.2 \\

ISR correction factor           &     1.5 \\
Mass window of $D_s^{\pm} \to K_S^0 K^{\pm}$   &     2.2 \\
Mass window of $D_s^{\pm} \to K^+ K^- \pi^{\pm}$   &     1.0 \\
Fit method                        & 6.4\\

\hline                                                                   
Total uncertainty                &     {\color{red}{12.4}}  \\
\hline
\end{tabular}
\label{tab:sys}
\end{center}
\end{table}

\section{Upper Limit With Systematic Uncertainty}

\begin{figure}[hbtp]
\centering

\begin{minipage}[b]{0.6\textwidth}
\epsfig{width=0.98\textwidth,clip=true,file=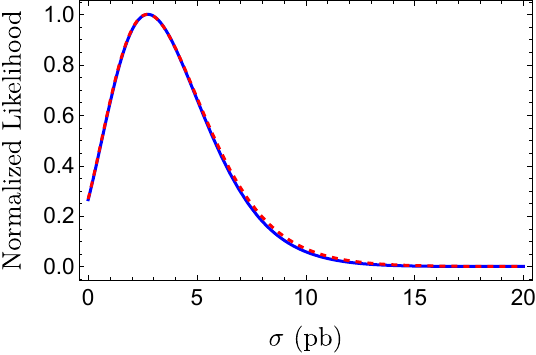}
\put(-120,110){\bf\boldmath $\sqrt{s}$ = 4.6988 GeV}
\end{minipage}
\caption{The likelihood distribution before (blue line) and after (red line) convolution.}
\label{appd}
\end{figure}

$~~~~$Since the statistical significance is only 1.2$\sigma$, no significant $e^+e^-\to \pi^+\pi^-D_s^+D_s^-$ signals are observed, an upper limit at the 90\% C.L. on the cross section is set using the Bayesian method. In this method, the probability distribution of the measured cross section $\sigma$, $P(\sigma)$, is determined using a maximum likelihood fit. The upper limit at the 90\% confidence level~(C.L.), denoted as $L$, is then calculated by solving the equation
\begin{linenomath*}
\begin{equation}
0.1 = \int_L^{\infty} P(\sigma) d\sigma.
\label{eq301}
\end{equation}
\end{linenomath*}
To include multiplicative systematic uncertainties from tracking, integrated luminosity, PID, branching fractions, and the ISR correction factor, $P(\sigma)$ is convolved with a probability distribution function of sensitivity, which refers to the denominator of Eq.~(\ref{eq111}) and is assumed to be a Gaussian with the central value $\hat{S}$ and the standard deviation $\sigma_s$~\cite{upp}. The convolution is performed as:
\begin{linenomath*}
\begin{equation}
P'(\sigma) = \int_0^{\infty} P\left(\frac{S}{\hat{S}}\sigma\right) \exp\left[-\frac{(S - \hat{S})^2}{2\sigma_s^2}\right] dS,
\label{eq302}
\end{equation}
\end{linenomath*}
where $P(\sigma)$ is the likelihood distribution obtained from fitting the likelihood of the number of signal events and parameterized as three Gaussians. The sensitivity is denoted by $\hat{S}$,  and its multiplicative systematic uncertainty is $\sigma_s$. Finally, the upper limit at the 90\% C.L. is obtained by integrating $P'(\sigma)$ rather than $P(\sigma)$ in Eq.~\eqref{eq301}.
Figure~\ref{appd} shows the likelihood distribution before and after convolution at the energy of 4.70~GeV. 

\section{Summary}
$~~~~$In this work, we report the study of the process $e^+e^-\to\pi^+\pi^-D_s^+D_s^-$ at c.m. energies from 4.42 to 4.95~GeV using BESIII data.
No significant signals are observed in this region. The Born cross sections are measured at fifteen energy values, and the corresponding upper limits at the 90\% C.L. are determined after considering the systematic uncertainties. These results are listed in TABLE~\ref{CROSS}. To further investigate the charm-strange quark dynamics and search for potential multi-quark structures, the $M(D_s^+ D_s^-)$ invariant mass spectrum was examined. No significant resonant structures are observed, with the present data set, underscoring the need for higher statistics in future investigations of this channel.

\begin{table}[h]
\begin{center}

\caption{The results of the Born cross sections for $e^+e^-\to \pi^+\pi^-D_s^+D_s^-$. The table includes the integrated luminosity $\mathcal{L}$, the number of signal events $N^{\rm obs}$, the ISR
correction factor $(1+\delta^{r})$ and the VP correction factor $(1+\delta^{v})$, the sum of the products of the branching fraction and eﬃciency $\sum_{i,j}\epsilon_{ij}\mathcal{B}_{i}\mathcal{B}_{j}$, and the Born cross section $\sigma$ along with its upper limit at the 90\% C.L. The first and second uncertainties are statistical and systematic.}
\begin{tabular}{c|ccccc}
\hline
$\sqrt{s}$ & $\mathcal{L}$ & $N^{\rm sig}$ & $\sum_{i,j}\epsilon_{ij}\mathcal{B}_{i}\mathcal{B}_{j}$ & $(1+\delta^{r})\times$ & $\sigma$ ($\sigma_{\rm UL}$) \\
(GeV) & (pb$^{-1}$) & & $(10^{-4})$ & $(1+\delta^{v})$ & (pb) \\
\hline
4.4156 & 1043.9 & $0.0^{+2.2}_{-0.0}$ & 2.871 & 0.863 & $0.0^{+8.2}_{-0.0}\pm0.0$ (<13.8) \\
4.4362 & 569.9  & $0.0^{+1.8}_{-0.0}$ & 2.933 & 0.896 & $0.0^{+11.5}_{-0.0}\pm0.0$ (<21.1) \\
4.5995 & 586.9  & $0.0^{+1.2}_{-0.0}$ & 3.296 & 0.975 & $0.0^{+4.9}_{-0.0}\pm0.0$ (<9.7) \\
4.6119 & 103.6  & $0.0^{+1.2}_{-0.0}$ & 2.796 & 0.995 & $0.0^{+31.5}_{-0.0}\pm0.0$ (<62.5) \\
4.6280 & 521.5  & $0.5^{+2.0}_{-0.3}$ & 2.738 & 1.001 & $2.6^{+10.5}_{-1.6}\pm0.3$ (<21.1) \\
4.6409 & 551.6  & $0.0^{+1.9}_{-0.0}$ & 2.812 & 1.008 & $0.0^{+9.4}_{-0.0}\pm0.0$ (<16.9) \\
4.6612 & 529.4  & $0.7^{+2.6}_{-0.4}$ & 2.795 & 1.014 & $3.9^{+14.3}_{-2.2}\pm0.5$ (<26.8) \\
4.6819 & 1667.4 & $1.0^{+3.4}_{-0.7}$ & 3.052 & 1.018 & $1.8^{+6.1}_{-1.2}\pm0.2$ (<11.3) \\
4.6988 & 535.5  & $2.2^{+3.0}_{-1.4}$ & 3.242 & 1.018 & $12.4^{+17.0}_{-7.9}\pm1.5$ (<39.6) \\
4.7397 & 163.9  & $0.0^{+1.4}_{-0.0}$ & 4.043 & 1.006 & $0.0^{+23.5}_{-0.0}\pm0.0$ (<46.9) \\
4.7501 & 366.5  & $0.0^{+1.5}_{-0.0}$ & 4.086 & 1.000 & $0.0^{+11.4}_{-0.0}\pm0.0$ (<22.7) \\
4.7805 & 511.5  & $0.5^{+2.8}_{-0.3}$ & 4.253 & 0.979 & $2.7^{+15.1}_{-1.6}\pm0.3$ (<28.0) \\
4.8431 & 525.2  & $0.4^{+2.8}_{-0.2}$ & 4.510 & 0.931 & $2.0^{+14.1}_{-1.0}\pm0.2$ (<25.1) \\
4.9180 & 207.8  & $0.0^{+1.4}_{-0.0}$ & 4.609 & 0.891 & $0.0^{+17.3}_{-0.0}\pm0.0$ (<34.7) \\
4.9509 & 159.3  & $2.2^{+2.8}_{-1.3}$ & 4.667 & 0.865 & $35.7^{+45.4}_{-21.1}\pm4.4$ (<108.9) \\
\hline
\end{tabular}
\label{CROSS}
\end{center}
\end{table}
\label{chap:SUMMARY}

\acknowledgments
The BESIII Collaboration thanks the staff of BEPCII (https://cstr.cn/31109.02.BEPC) and the IHEP computing center for their strong support. This work is supported in part by National Key R\&D Program of China under Contracts Nos. 2023YFA1606704, 2025YFA1613900, 2023YFA1606000; National Natural Science Foundation of China (NSFC) under Contracts Nos. 11635010, 11935015, 11935016, 11935018, 12025502, 12035009, 12035013, 12061131003, 12192260, 12192261, 12192262, 12192263, 12192264, 12192265, 12221005, 12225509, 12235017, 12342502, 12361141819, 12535005, 12575089; the Chinese Academy of Sciences (CAS) Large-Scale Scientific Facility Program; the Strategic Priority Research Program of Chinese Academy of Sciences under Contract No. XDA0480600; CAS under Contract No. YSBR-101; 100 Talents Program of CAS; The Institute of Nuclear and Particle Physics (INPAC) and Shanghai Key Laboratory for Particle Physics and Cosmology; ERC under Contract No. 758462; German Research Foundation DFG under Contract No. FOR5327; Istituto Nazionale di Fisica Nucleare, Italy; Knut and Alice Wallenberg Foundation under Contracts Nos. 2021.0174, 2021.0299, 2023.0315; Ministry of Development of Turkey under Contract No. DPT2006K-120470; National Research Foundation of Korea under Contract No. NRF-2022R1A2C1092335; National Science and Technology fund of Mongolia; Polish National Science Centre under Contract No. 2024/53/B/ST2/00975; STFC (United Kingdom); Swedish Research Council under Contract No. 2019.04595; U. S. Department of Energy under Contract No. DE-FG02-05ER41374.



\end{document}

%% file: author.tex
M.~Ablikim$^{1}$\BESIIIorcid{0000-0002-3935-619X},
M.~N.~Achasov$^{4,c}$\BESIIIorcid{0000-0002-9400-8622},
P.~Adlarson$^{82}$\BESIIIorcid{0000-0001-6280-3851},
X.~C.~Ai$^{88}$\BESIIIorcid{0000-0003-3856-2415},
C.~S.~Akondi$^{31A,31B}$\BESIIIorcid{0000-0001-6303-5217},
R.~Aliberti$^{39}$\BESIIIorcid{0000-0003-3500-4012},
A.~Amoroso$^{81A,81C}$\BESIIIorcid{0000-0002-3095-8610},
Q.~An$^{78,65,\dagger}$,
Y.~H.~An$^{88}$\BESIIIorcid{0009-0008-3419-0849},
M.~S.~Anderson$^{39}$\BESIIIorcid{0009-0008-1550-2632}
% ...
\\
\vspace{0.2cm}
(BESIII Collaboration)\\
\vspace{0.2cm} {\it
$^{1}$ Institute of High Energy Physics, Beijing 100049, People's Republic of China\\
% ...
$^{4}$ Budker Institute of Nuclear Physics SB RAS (BINP), Novosibirsk 630090, Russia\\
% ...
$^{31}$ INFN Sezione di Ferrara, (A)INFN Sezione di Ferrara, I-44122, Ferrara, Italy; (B)University of Ferrara, I-44122, Ferrara, Italy\\
% ...
$^{39}$ Johannes Gutenberg University of Mainz, Johann-Joachim-Becher-Weg 45, D-55099 Mainz, Germany\\
% ...
$^{65}$ State Key Laboratory of Particle Detection and Electronics, Beijing 100049, Hefei 230026, People's Republic of China\\
% ...
$^{78}$ University of Science and Technology of China, Hefei 230026, People's Republic of China\\
% ...
$^{81}$ University of Turin and INFN, (A)University of Turin, I-10125, Turin, Italy; (B)University of Eastern Piedmont, I-15121, Alessandria, Italy; (C)INFN, I-10125, Turin, Italy\\
$^{82}$ Uppsala University, Box 516, SE-75120 Uppsala, Sweden\\
% ...
$^{88}$ Zhengzhou University, Zhengzhou 450001, People's Republic of China\\

\vspace{0.2cm}
$^{\dagger}$ Deceased\\
% ...
$^{c}$ Also at the Novosibirsk State University, Novosibirsk, 630090, Russia\\
% ...

}